\documentclass[a4paper, amsfonts, amssymb, amsmath, preprint, showkeys, nofootinbib, reprint,superscriptaddress]{revtex4-1}
\usepackage[utf8]{inputenc}
\usepackage{bm}
\usepackage[colorinlistoftodos, color=green!40, prependcaption]{todonotes}
\usepackage{amsthm}
\usepackage{mathtools}
\usepackage{physics}
\usepackage{xcolor}
\usepackage{graphicx}
\usepackage[left=23mm,right=13mm,top=35mm,columnsep=15pt]{geometry} 
\usepackage{adjustbox}
\usepackage{placeins}
\usepackage[T1]{fontenc}
\usepackage{lipsum}
\usepackage{csquotes}
\usepackage[pdftex, pdftitle={Article}, pdfauthor={Author}]{hyperref} 
\newcommand{\ten}{\mathsf{T}}

\newcommand{\EE}{\mathbb{E}}
\newcommand{\tpr}{{t^\prime }}

\newcommand{\sfF}{\mathsf{F}}
\newcommand{\GammaCrit}{{\Gamma_\beta^\star}}

\newcommand{\hchi}{\hat{\chi}}
\newcommand{\hq}{\hat{q}}
\newcommand{\hQ}{\hat{Q}}
\newcommand{\ham}{\hat{m}}

\newcommand{\av}[1]{{\langle #1 \rangle}}

\newcommand{\eff}{{\rm eff}}
\newcommand{\zsf}{\mathsf{z}}
\DeclareMathOperator*{\Extr}{{\rm Extr}}
\newcommand*{\rev}{\textcolor{black}}

\begin{document}
\title{Exact Replica Symmetric Solution for Transverse Field Hopfield Model under Finite Trotter Size}

\author{Koki Okajima}
    \email[Correspondence email address: ]{darjeeling@g.ecc.u-tokyo.ac.jp}
    \affiliation{Department of Physics, Graduate School of Science, The University of Tokyo, 7-3-1 Hongo, Bunkyo-ku, Tokyo 113-0033, Japan}
\author{Yoshiyuki Kabashima}
    \affiliation{Department of Physics, Graduate School of Science, The University of Tokyo, 7-3-1 Hongo, Bunkyo-ku, Tokyo 113-0033, Japan}
    \affiliation{Institute for Physics of Intelligence, Graduate School of Science,
The University of Tokyo, 7-3-1 Hongo, Bunkyo-ku, Tokyo 113-0033, Japan.}
\affiliation{Trans-Scale Quantum Science Institute, The University of Tokyo, 7-3-1 Hongo, Bunkyo-ku, Tokyo 113-0033, Japan}

\date{\today} 

\begin{abstract}

We analyze the quantum Hopfield model in which an extensive number of patterns are embedded in the presence of a uniform transverse field. This analysis employs the replica method  under the replica symmetric ansatz on the Suzuki-Trotter representation of the model, while keeping the number of Trotter slices $M$ finite. The statistical properties of the quantum Hopfield model in imaginary time are reduced to an effective $M$-spin long-range classical Ising model, which can be extensively studied using a dedicated Monte Carlo algorithm. This approach contrasts with the commonly applied static approximation, which ignores the imaginary time dependency of the order parameters, but allows $M \to \infty$ to be taken analytically. During the analysis, we introduce an exact but fundamentally weaker static relation, referred to as the quasi-static relation. We present the phase diagram of the model with respect to the transverse field strength and the number of embedded patterns, indicating a small but quantitative difference from previous results obtained using the static approximation.
\end{abstract}


\maketitle

\section{Introduction} \label{sec:outline}
The Hopfield model \cite{hopfieldNeuralNetworksPhysical1982} has been instrumental in the study of artificial neural networks and associative memory. Research using statistical mechanical techniques, such as the replica method \cite{RSB40}, has provided critical insights into the model's properties, including its storage capacity and phase diagram \cite{amitSpinglassModelsNeural1985, amitStoringInfiniteNumbers1985}. In recent years, developments in machine learning have introduced mechanisms analogous to the Hopfield model. 
Examples include the modern Hopfield network for pattern recognition \cite{krotovDenseAssociativeMemory2016, theriaultDenseHopfieldNetworks2024, demircigilModelAssociativeMemory2017}, the Attention mechanism in Transformers \cite{ramsauerHopfieldNetworksAll2020}, and Factorization machines \cite{camilliMatrixFactorizationNeural2023, Camilli_2024}. These advancements have prompted a renewed examination of the Hopfield model in contemporary contexts.

Building on the interest in the Hopfield model as an information processing system, it is worthwhile to explore the effect of integrating quantum effects into the model, particularly in the context of quantum optimization methods such as \textit{quantum annealing} (QA)\cite{kadowakiQuantumAnnealingTransverse1998, santoroOptimizationUsingQuantum2006}. QA aims to find the ground state of a spin system via quantum fluctuations, in contrast to the more popular \textit{simulated annealing}, which uses thermal fluctuations \cite{SimulatedAnnealing}. 

The Hopfield model under quantum effects has been previously investigated using the Trotter decomposition in conjunction with the replica method \cite{nishimoriQuantumEffectsNeural1996, sekiQuantumAnnealingAntiferromagnetic2015}. In this formulation, an exact treatment requires taking the number of Trotter slices to infinity, which is unfortunately difficult to achieve analytically. To overcome this difficulty, a simplification known as the static approximation (SA) \cite{brayReplicaTheoryQuantum1980} is employed. This approximation assumes that the order parameters of the system lose their imaginary time dependency. There are significant advantages to employing SA; it allows the number of Trotter slices to be taken to infinity analytically, eliminating the effects of finite Trotter size. Additionally, it makes it possible to analytically reach the zero-temperature limit, enabling a pure investigation of the quantum effects on the model without thermal fluctuations.

Although SA is analytically tractable and can provide simple formulae that are provably accurate in special cases \cite{okuyamaExactSolutionPartition2018}, its overall accuracy has been scarcely studied. Research on systems without employing SA has been limited to spin glasses in the paramagnetic phase \cite{usadelQuantumFluctuationsIsing1987, buttnerStabilityAnalysisIsing1990, youngStabilityQuantumSherringtonKirkpatrick2017}, or to models under random quantum fields \cite{shiraishi2024exactlysolvablequantumspinglass}, where the free energy can be obtained in a mathematically rigorous manner. 
\rev{Effective theories beyond SA have also been proposed by splitting the order parameters into static and dynamic components, in which the contribution from the latter to a renormalized free energy is handled approximately via Landau expansion \cite{KazutakaTakahashi07}. }
Only recently have the low-temperature properties of the quantum Sherrington-Kirkpatrick \cite{kissCompleteReplicaSolution2024} and Heisenberg
 \cite{kavokineExactNumericalSolution2024} models been investigated without SA, revealing that predictions by SA can significantly deviate from the correct values. This motivates a critical reconsideration of SA for other models.
 
This paper is based on these considerations. Specifically, to examine the accuracy of SA, we analyze a quantum Hopfield model, where the quantum effect is introduced by a uniform transverse field, using the replica method without employing SA. Although SA allows for analytically taking the number of Trotter slices to infinity, its solution generally does not satisfy the self-consistent equation derived by the replica method. In contrast, our approach constructs solutions that satisfy the self-consistent equation under an assumed ansatz concerning replica symmetry while keeping the number of Trotter slices finite. The accuracy of SA is examined by comparing it with the extrapolation of the constructed solution to the limit of an infinite number of Trotter slices, either numerically or analytically in special cases.

This paper is organized as follows. Section \ref{sec:RS_solution} explains the main results of the analysis under finite Trotter size. In this section, we introduce the quasi-static ansatz, a weaker version of SA, which we find to be exact over a broad range of the phase diagram. Building upon this analysis, Section \ref{sec:alpha_zero_Crit} provides observations on the quantum phase transition point of the model under a small number of patterns, suggesting that the point predicted by SA is consistent with our analysis up to first order. In Section \ref{sec:numerics}, we present a numerical analysis of the Hopfield model and its phase diagram for finite Trotter size, extrapolating the obtained solution to gain insight into the limit of an infinite number of Trotter slices. We also discuss some implications on the accuracy of SA.  
\section{ Replica Symmetric Analysis of the Transverse Field Hopfield Model } \label{sec:RS_solution}
The Hamiltonian of the two-body Hopfield model with $N$ spins is defined by 
\begin{equation}
    -\hat{\mathcal{H}}_0 = \frac{1}{N} \sum_{i < j}  \sum_{\mu = 1} ^P \xi_i^\mu \xi_j^\mu  \hat{\sigma}^z_i \hat{\sigma}^z_j,
\end{equation}
where $\hat{\sigma}_i^{x,y,z}$ represent the Pauli spin matrices, and 
$\bm{\xi}^1, \bm{\xi}^2, \ldots, \bm{\xi}^P \in \{-1,+1\}^N$ are the $P$  binary patterns embedded in the Hopfield model. {We consider the non-trivial case where $P$ is proportional to $N$, \textit{i.e.} $P = \alpha N$ for $\alpha = O(1)$, since SA is known to be exact when $P$ is sublinear in $N$ \cite{okuyamaExactSolutionPartition2018}.}
Additionally, we consider the transverse field with intensity $\Gamma > 0$, in which case the total Hamiltonian is given by 
\begin{equation}
    -\hat{\mathcal{H}} = -\hat{\mathcal{H}}_0 + \Gamma \sum_{i = 1}^N \hat{\sigma}_i^x. 
\end{equation}
The partition function written in the Suzuki-Trotter formulation \cite{Trotter1959, Suzuki1976} is given by 
\begin{equation}
    \mathcal{Z} = \lim_{M\to \infty} Z_M, 
\end{equation}
{where $Z_M$ is the partition function of a classical spin model consisting of $NM$ number of Ising spins $\{\sigma_{i,t}\} \in \{-1,+1\}^{NM}$, indexed by site coordinate $i = 1, \ldots, N$ and imaginary time coordinate $t = 0, \ldots, M-1$: }
\begin{align}\label{eq:Trotter_Partition}
\begin{split}
        {Z}_M \equiv \mathop{\Tr}_{\{\sigma_{i,t}\}  } \exp \Bigg[ \frac{\eta}{N} \sum_{t = 0}^{M-1} \sum_{i < j} \sum_{\mu = 1}^P \xi_i^\mu \xi_j^\mu  \sigma_{i,t} \sigma_{j,t} \\
        \qquad \qquad \qquad + B \sum_{t = 0}^{M-1}  \sum_{i = 1}^N \sigma_{i,t} \sigma_{i,t+1}  \Bigg].
\end{split}
\end{align}
Here, $\eta \equiv \beta / M$,  $B \equiv -\frac{1}{2} \log \tanh \eta \Gamma $, and a periodic boundary condition with respect to imaginary time is imposed via 
$\sigma_{i,M} \equiv \sigma_{i,0}$ for all $1 \leq i \leq N$. 
Note that the imaginary time is indexed from zero, which we will find convenient in the analysis that follows in this section. 
Assuming that the embedded patterns consist of independent and identically distributed entries, the average of the log partition function for finite $M$ under infinite $N$ can be calculated by a standard replica calculation \cite{FedorovQuantumSpinGlass1986, usadelQuantumFluctuationsIsing1987, buttnerStabilityAnalysisIsing1990}; see Appendix \ref{sec:RS_FE_derivation} for a detailed derivation. Here, we consider the case where the spins only have a non-vanishing overlap with a single pattern $\bm{\xi}^1$, which we take as $\bm{\xi}^1 = (1, \ldots, 1)^\ten = \bm{1}$ without loss of generality. 
Under the replica symmetric assumption, the free energy per spin up to a constant in the thermodynamic limit is given by the extremum of a function of order parameters, which consist of $M\times M$ symmetric semi-positive definite matrices and a $M$-dimensional vector:
\begin{widetext}
\begin{align}
\label{eq:replica_symmetric_free_energy}
\begin{split}
- \lim_{N\to \infty} \frac{1}{N\beta} \EE \log Z_M = &\Extr_{ \substack{\bm{q}, \bm{\hq}, \bm{\chi}, \bm{\hchi} \in \mathbb{S}_+^M, \\  \bm{m} \in \mathbb{R}^M } } \Bigg\{ \frac{ \Tr \bm{q}\bm{\hq}}{2M}  +\frac{\Tr \bm{\chi} \bm{\hchi} }{2M} + \frac{\norm{\bm{m}}_2^2}{2M}  + \frac{\alpha}{2\beta} \log \det \qty(\bm{I}_M - \eta \bm{\chi}) \\
    &- \frac{\alpha}{2M} \Tr (\bm{I}_M - \eta\bm{\chi})^{-1}  \qty(\bm{q} - {\bm{\chi}}) - \frac{1}{\beta} \EE_{\bm{z} \sim \mathcal{N}(\bm{0}, \bm{\hchi} / \eta )} \log \mathop{\Tr}_{\{\sigma_t\}} e^{-H_{\rm eff}}  \Bigg\} ,
    \end{split}
\end{align}
where $\mathbb{S}_+^M = \{ \bm{X} \in \mathbb{R}^{M \times M} \ | \ \bm{X} = \bm{X}^{\mathsf{T}}, \ \bm{X} \succeq 0 \}$, and $\bm{I}_M$ is the identity matrix of size $M \times M$. 
The effective Hamiltonian $H_{\rm eff}$ is given by a long-range Ising model under a periodic boundary condition: 
\begin{equation}\label{eq:Effective_Hamiltonian}
        -H_{\rm eff}
        = \frac{\eta}{2} \sum_{t, \tpr} \hq_{t\tpr} \sigma_t \sigma_\tpr + \eta  \sum_{t = 0}^{M-1} (z_t +m_t ) \sigma_{t} + B \sum_{t = 0}^{M-1} \sigma_t \sigma_{t+1}.
\end{equation}
\end{widetext}

Let us elaborate on the physical meaning of the order parameters introduced earlier. 
Both matrix order parameters $\bm{q} = (q_{t\tpr})$ and $\bm{\chi} = (\chi_{t\tpr})$ represent the average of two types of two-point correlation functions between different imaginary times with respect to $\{\bm{\xi}^\mu\}$, given elementwise by: 
\begin{align}
    q_{t\tpr} & = \frac{1}{N}\EE \Bigg[ \sum_{i = 1}^N \av{\sigma_{i,t} \sigma_{i,\tpr}} \Bigg],\\
    \chi_{t\tpr} & =\frac{1}{N} \EE  \Bigg[\sum_{i = 1}^N \av{\sigma_{i,t} \sigma_{i,\tpr}} - \av{\sigma_{i,t}}\av{ \sigma_{i,\tpr}} \Bigg],
\end{align}
where $\av{\cdots}$ represents the thermal average with respect to the original thermodynamical system, given by \eqref{eq:Trotter_Partition}. Consequently, the extremum conditions with respect to $\bm{q}$ and $\bm{\chi}$ offers 
\begin{align}\label{eq:q_extr}
    q_{t\tpr} &=  \EE_{\bm{z} \sim \mathcal{N}(\bm{0}, \bm{\hchi} / \eta )} \Big[ \av{\sigma_t \sigma_\tpr}_{\rm eff} \Big], \\
    \label{eq:chi_extr}
    \chi_{t\tpr} &= \EE_{\bm{z} \sim \mathcal{N}(\bm{0}, \bm{\hchi} / \eta )} \Big[ \av{\sigma_t \sigma_\tpr}_{\rm eff} - \av{\sigma_t}_{\rm eff} \av{ \sigma_\tpr}_{\rm eff} \Big], 
\end{align}
where $\av{\cdots}_{\rm eff}$ represents the thermal average with respect to the effective model \eqref{eq:Effective_Hamiltonian}. The same observation can be made on the vector order parameter $\bm{m} = (m_t)$, in which case they express the overlap with the retrieved pattern $\bm{\xi}^1$ (note that we take $\xi_i^1 = 1$ for all $i$):
\begin{align}
    m_t &= \frac{1}{N} \EE \Bigg[ \sum_{i = 1}^N \av{\sigma_{i,t}} \Bigg] =  \EE_{\bm{z} \sim \mathcal{N}(\bm{0}, \bm{\hchi} / \eta )} [\av{\sigma_t}].
\end{align}
The matrices $\bm{\hq}$ and $ \bm{\hchi}$ are the conjugate matrices of $\bm{q}$ and $\bm{\chi}$. 
\rev{
A direct derivative of the free energy with respect to $\bm{q}$ yields $\bm{\hq} = \alpha (\bm{I}_M - \eta \bm{\chi})^{-1}$. However, since the diagonals of $\bm{q}$ are fixed to unity,
for $a \in \mathbb{R}$, the transformation $\bm{\hat{q}} \to \bm{\hat{q}} + a \bm{I}_M$ only induces constant shifts in both the free energy and the effective Hamiltonian. This allows us to choose $a$ arbitrarily. 
Taking this into consideration, the order parameter $\bm{\hq}$ can be given by
\begin{equation}\label{eq:hatq_extr}
    \bm{\hq} = \alpha \eta \bm{\chi} (\bm{I}_M - \eta \bm{\chi})^{-1},
\end{equation}
with the choice $a = \alpha \eta$. 
The extremum condition for $\bm{\hchi}$ is given by 
\begin{equation}
    \label{eq:hatchi_extr}
    \bm{\hchi}= \alpha \eta (\bm{I}_M - \eta \bm{\chi})^{-1}(\bm{q} - \bm{\chi}) (\bm{I}_M - \eta \bm{\chi})^{-1}.
\end{equation}
}
The specific profile of the spin interactions and fields in the effective Hamiltonian is determined by solving (\ref{eq:q_extr} - \ref{eq:hatchi_extr}), which we refer to as the equations of state, in a self-consistent manner.

\subsection{Circulant property of the order parameters and the quasi-static ansatz (qSA)}
It is not difficult to notice that, if the long-range interactions and the covariance of the external field of the effective Hamiltonian only depend on the distance between two spins, \textit{i.e.} $\hq_{t\tpr} = \hq_{\abs{t-\tpr}}$ and $\hchi_{t\tpr} = \hchi_{\abs{t-\tpr}},$ then the same properties hold for the two-point correlation functions, \textit{i.e.} $q_{t\tpr} = q_{\abs{t-\tpr}}$ and $\chi_{t\tpr} = \chi_{\abs{t-\tpr}}$, with the absolute value distance being defined on a ring. Such matrices are referred to as symmetric \textit{circulant} matrices, which are simultaneously diagonalizable using the complex discrete Fourier matrix of size $M$, $\sfF_M$ \cite{CirculantMatrices06}: 
\begin{align}
\label{eq:FourierRep_q}
    \bm{q} &= \sfF_M  {\rm diag}(q_0, \ldots, q_{M-1})  \sfF^\dagger_M,\\
    \label{eq:FourierRep_chi}
    \bm{\chi} &= \sfF_M  {\rm diag}(\chi_0, \ldots, \chi_{M-1})  \sfF^\dagger_M, 
\end{align}
where $q_k \in \mathbb{R}$ and $ \chi_k \in \mathbb{R}$ are the eigenvalues of $\bm{q}$ and $\bm{\chi}$ corresponding to the $k$-th frequency component, or $k$-th mode in discrete Fourier space. Note that the zeroth mode corresponds to the constant or uniform component, 
where $(\sfF_M)_{t,0} = 1/\sqrt{M}$ for $t = 0, \ldots, M-1$. From this circulant nature of the effective Hamiltonian, the magnetization $\bm{m}$ does not possess imaginary time dependency, \textit{i.e.} $m_t = m$ for $t = 0, \ldots, M-1$. 

Let us pay particular attention to the external field of the effective Hamiltonian, whose imaginary time dependency only arises from the random term $\bm{z} \sim \mathcal{N}(\bm{0}, \bm{\hchi} / \eta)$. If $\bm{\hchi}$ only possesses the zeroth mode, \textit{i.e.} $\bm{\hchi} = \hchi_0 \bm{1}\bm{1}^\mathsf{T} / M$, then $z_t = \mathsf{z}$ for $t = 0,\ldots, M-1$, where $\mathsf{z} \sim \mathcal{N}(0, \hchi_0 / \beta)$. Under a uniform external field, $\av{\sigma_t}_{\rm eff}$ loses its imaginary time dependency for any realization of random variable $\mathsf{z}$. Substituting \eqref{eq:q_extr} and \eqref{eq:chi_extr} to \eqref{eq:hatchi_extr}, we find 
\begin{align}\label{eq:qSA_chihat}
    \bm{\hchi} &=  \alpha \beta \frac{\EE_{\mathsf{z}} [ \av{\sigma_1}_{\rm eff}^2 ]  }{(1 - \eta \chi_0)^2} \frac{\bm{1}\bm{1}^\ten}{M}
\end{align}
where $\EE_{\mathsf{z}}$ denotes the average with respect to $z_0 = \ldots = z_{M-1} = \mathsf{z} \sim \mathcal{N}(0, \hchi_0 / \beta)$. { This indicates that there may exist a fixed point of the self-consistent equations (\ref{eq:q_extr}-\ref{eq:hatchi_extr}) where $\bm{\hchi}$ is a uniform matrix. }
We refer to such solutions as \textit{quasi-static}, and the assumption of such property as the \textit{quasi-static ansatz} (qSA). 
\rev{Such properties have already been suggested in preliminary works such as \cite{KazutakaTakahashi07} and \cite{kissCompleteReplicaSolution2024}. }
This is in contrast with the conventional SA, which further assumes that $\bm{\hq}$ is a uniform matrix.{  Note that SA is clearly false since the two-point correlation function under the effective model \eqref{eq:chi_extr} cannot be uniform under any uniform long-range interaction and external field of finite quantity, and thus the right-hand side of \eqref{eq:hatq_extr} cannot be a uniform matrix. The result from SA can be recovered by truncating the matrix $\bm{\hq}$ to a single rank by projecting the right-hand side of \eqref{eq:hatq_extr} to the subspace in $\mathbb{R}^M$ spanned by $\bm{1}$:}
\begin{equation} \label{eq:SA_qhat}
    \text{(SA)} \quad \bm{\hq} =  \alpha \eta \mathcal{P}_{\bm{1}} [ \bm{\chi} (\bm{I}_M - \eta \bm{\chi})^{-1} ] =\frac{  \alpha \eta \chi_0}{1 - \eta \chi_0} \frac{\bm{1}\bm{1}^\ten}{M},
\end{equation}
where $\mathcal{P}_{\bm{1}}$ is the projection operator under consideration. 
Therefore, a clear distinction should be made between the two; while qSA can potentially be exact, SA only serves as an approximation excluding special cases \cite{okuyamaExactSolutionPartition2018}. 

\subsection{Stability of the quasi-static solution}

The stability of a quasi-static solution can be determined by a standard perturbation analysis.  Consider the perturbed matrix $\bm{\hchi}^\prime$ given by 
\begin{equation}
    \bm{\hchi}^\prime = \hchi_0 \frac{\bm{1} \bm{1}^\mathsf{T}}{M} + \delta \bm{\hchi},
\end{equation}
where the perturbation matrix $\delta \bm{\hchi}$ is circulant-symmetric, with its null space spanned by $\bm{1}$. In other words, its discrete Fourier representation is 
\begin{equation}
    \delta \bm{\hchi} = \sfF_M {\rm diag} ( 0, \delta \hchi_1, \ldots, \delta \hchi_{M-1}) \sfF_M^\dagger.
\end{equation}
Expanding $\bm{q} - \bm{\chi}$ up to first order of $\delta \bm{\hchi}$, we find 
\begin{equation}
\begin{gathered}
    \EE_{\bm{z} \sim \mathcal{N} (\bm{0} , \bm{\hchi}^\prime  / \eta ) }[  \av{\bm{\sigma}}_{\rm eff} \av{\bm{\sigma}}_{\rm eff}^{\mathsf{T}} ] - \EE_{\mathsf{z}} [ \av{\sigma_1}_{\rm eff}^2 ] \frac{\bm{1}\bm{1}^\ten}{M} \\
    = \frac{1}{2\eta}  \EE_{\mathsf{z}} \Big[ (\nabla^\ten \delta \bm{\hchi} \nabla)  \av{\bm{\sigma}}_{\rm eff} \av{\bm{\sigma}}_{\rm eff}^{\mathsf{T}} \Big] + O ( \| \delta \bm{\hchi} \|^2  ), 
\end{gathered}
\end{equation}
\rev{
where we used the differential operator expression of Gaussian integrals and a Taylor expansion with respect to $\delta \bm{\hchi}$, \textit{i.e.} 
\begin{equation}
    \begin{gathered}
        \EE_{\bm{z} \sim \mathcal{N} (\bm{0} , \bm{\hchi}^\prime  / \eta ) } [f(\bm{z})]
        =  \eval{e^{\frac{1}{2\eta} \nabla^\ten \bm{\hchi}^\prime \nabla } f(\bm{z}) }_{\bm{z} = \bm{0}} \\
        =  \eval{ e^{\frac{\hchi_0}{2\beta} (\nabla^\ten \bm{1})^2} \Big( 1 + \frac{\nabla^\ten \delta \bm{\hchi}\nabla}{2\eta} + O( \|  \delta \bm{\hchi} \|^2)  \Big)  f(\bm{z}) }_{\bm{z} = \bm{0}} 
    \end{gathered}
\end{equation}
in which the exponential differential operator is reverted back to its Gaussian average representation.
}
Using the equality $\pdv{}{z_t} \av{\mathcal{O}}_{\rm eff} = \av{\mathcal{O} \sigma_t}_{\rm eff} - \av{\mathcal{O}}_{\rm eff} \av{\sigma_t}_{\rm eff} $, and taking note that the null space of $\delta \bm{\hchi}$ is spanned by $\bm{1}$, the first order term is given by 
\begin{equation}
    \frac{1}{2\eta} \EE_{\mathsf{z}} \Bigg[ (\nabla^\ten \delta \bm{\hchi} \nabla)  \av{\bm{\sigma}}_{\rm eff} \av{\bm{\sigma}}_{\rm eff}^{\mathsf{T}} \Bigg] = \eta \EE_{\mathsf{z}} [ \bm{C}(\zsf) \delta \bm{\hchi} \bm{C}^\ten (\zsf) ],
\end{equation}
where $\bm{C}(\zsf) \in \mathbb{R}^{M\times M}$ is given by $C_{t\tpr}(\zsf) = \av{\sigma_t \sigma_\tpr}_{\rm eff} - \av{\sigma_t}_{\rm eff} \av{\sigma_\tpr}_{\rm eff}$ elementwise. Note that for any fixed $\mathsf{z}$, $\bm{C}(\zsf)$ is a symmetric circulant matrix, and hence share the same basis with $\delta \bm{\hchi}$ and $\bm{\chi}$. Therefore, by inserting this expression into \eqref{eq:hatchi_extr}, we have in terms of eigenvalues 
\begin{equation}
    \delta \chi_k = \alpha \eta^2 \frac{ \EE_{\mathsf{z}} [ C_k^2(\zsf)] }{  ( 1 - \eta \chi_k )^2 } \delta \chi_k + O( \| \delta \bm{\hchi} \|^2 ),
\end{equation}
for $k  \geq 1$. The quasi-static solution is stable if the above equation is stable around $\delta \chi_k = 0$ for all $k \geq 1$, which offers the condition 
\begin{equation} \label{eq:qSA_stability}
   \forall k \geq 1, \quad \alpha \eta^2 \frac{ \EE_{\mathsf{z}} [ C_k^2(\zsf)] }{  ( 1 - \eta \chi_k )^2 } < 1.
\end{equation}

The local stability condition of the replica symmetric solution, which correponds to the de Almeida-Thouless condition for classical spin glasses \cite{AT1978}, turns out to be given by the same condition as \eqref{eq:qSA_stability} for the zeroth mode \cite{HaraKabashima24} : 
\begin{equation} \label{eq:RS_stability}
     \alpha \eta^2 \frac{ \EE_{\mathsf{z}} [ C_0^2(\zsf)] }{  ( 1 - \eta \chi_0 )^2 } < 1,
\end{equation}
where $ C_0(\zsf)$ can be written alternatively as
\begin{equation}
    C_0(\zsf) = \frac{1}{M}\sum_{t,\tpr}\av{\sigma_t \sigma_\tpr}_{\rm eff} - \frac{1}{M} \qty( \sum_{t = 0}^{M-1} \av{\sigma_t}_{\rm eff} )^2.
\end{equation}
The matrix $\bm{C}(\zsf)$ is not only positive definite but also possesses positive entries due to Griffith's second inequality \cite{Griffiths67}. 
Therefore, the largest eigenvalue of $\bm{C}(\zsf)$ is given by $C_0(\zsf)$ from the Perron-Frobenius theorem. The same argument holds for $\bm{\chi} = \EE_{\mathsf{z}} [ \bm{C}(\zsf)]$, whose largest eigenvalue is $\chi_0$. This indicates that \eqref{eq:RS_stability} suffices for \eqref{eq:qSA_stability} to hold; in other words, it is equally reasonable to also assume qSA when replica symmetry is considered. 
In fact, for all numerical experiments in Section \ref{sec:numerics}, we found that \eqref{eq:qSA_stability} does indeed hold, demonstrating the robustness of qSA. 

\section{Critical Behavior near $\alpha = 0$}
\label{sec:alpha_zero_Crit}
The phase diagram of the system under any $\alpha, M, \Gamma$ and $\beta$ can be obtained by solving the equations of state numerically. From the perspective of pattern retrieval, we are interested in the critical properties of the retrieval phase, where a spontaneous magnetization $m > 0$ appears. For the quantum Hopfield model, there are known to be two types of phase transitions; one where the retrieval state becomes locally stable (R-I), and the other where the retrieval state becomes globally stable (R-II). Under small $\alpha$, the behavior of these transitions with respect to $\Gamma$ for finite $\beta$ can be obtained analytically via a perturbation analysis. Surprisingly, at least up to the first order of $\alpha$, one can also take the limit $M \to \infty$ in an exact manner, as we demonstrate that the zeroth mode of each matrix order parameter turns out to be the only relevant one.

Under small $\alpha$, the R-I critical point can be obtained from the condition that the equation 
\begin{equation}
    m = \EE_{\mathsf{z}} [ \av{\sigma_1}_{\rm eff} ]
\end{equation}
holds a non-zero solution. Note that when $\alpha$ is strictly zero, this condition reduces to 
\begin{equation}\label{eq:alpha0_crit}
    \Gamma = \tanh \beta \Gamma, 
\end{equation}
which we denote its solution as $\GammaCrit(\alpha = 0),$ as a special case of the critical transverse field for arbitrary $\alpha$, $\GammaCrit(\alpha)$. The objective of our analysis is then the function
\begin{equation}
    \Delta \Gamma_\beta(\alpha) \equiv \GammaCrit(0) - \GammaCrit(\alpha),
\end{equation}
which we aim to study up to its leading order of $\alpha$ in this section. 

\begin{figure*}[t]
    \centering
    \includegraphics[width=0.85\linewidth]{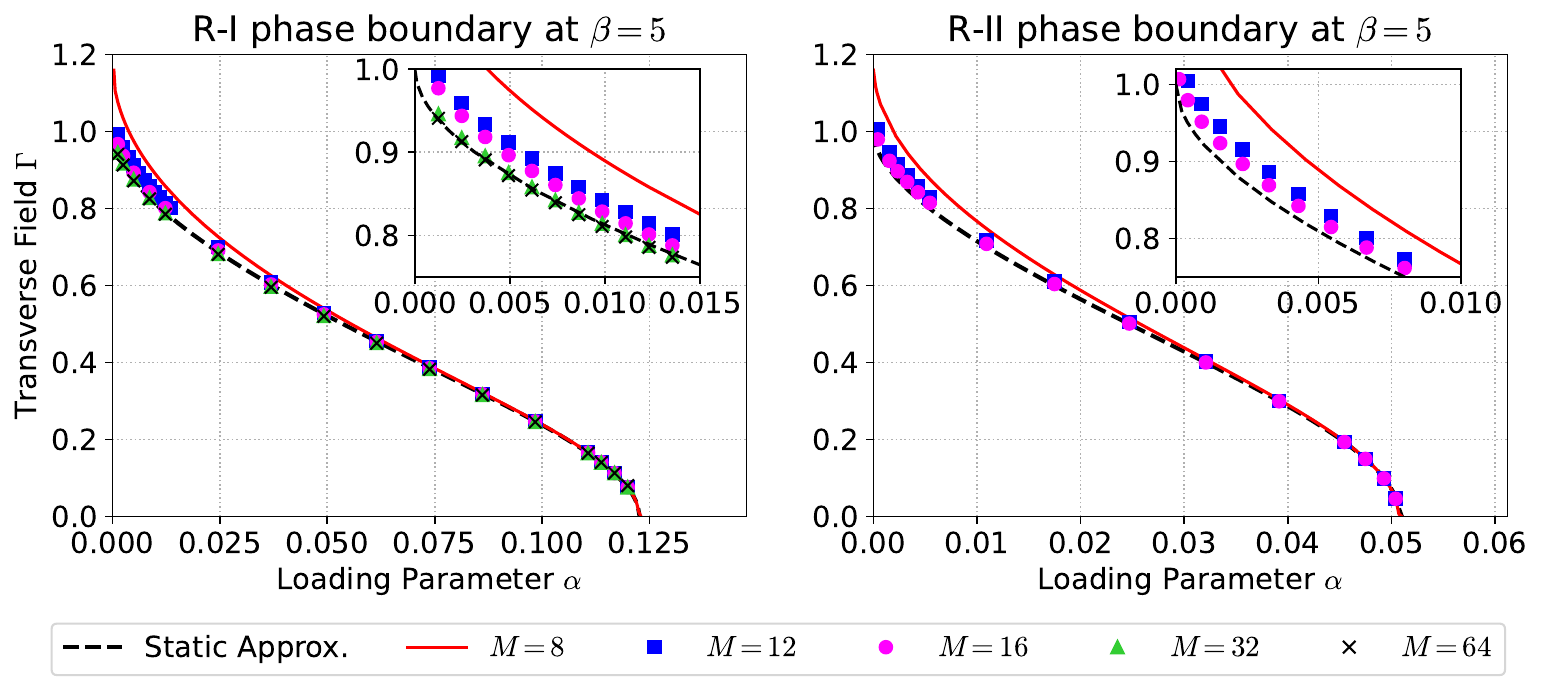}
    \caption{(Color online) The phase diagram of the Hopfield model under transverse field with $\beta = 5$ for different values of Trotter size $M$. }
    \label{fig:PhaseDiagram_Beta5}
\end{figure*}

Assuming that $m$, $\bm{\hq}$ and $\bm{\hchi}$ are all small, only the leading order of $\bm{\hq}$ and $\bm{\hchi}$ for $\alpha$ is relevant to the first-order analysis. Under this hypothesis, the leading order of the right-hand side of \eqref{eq:chi_extr} is given by the two-point correlation function of an Ising model with nearest-neighbor interaction of strength $B$ and external field $m$, which is given by 
\begin{equation}
\label{eq:Correlation}
\begin{gathered}
      \av{\sigma_t \sigma_\tpr} - \av{\sigma_t} \av{\sigma_\tpr}  \\ =\frac{m^2}{H^2} \sech^2\beta H + \frac{\Gamma^2}{H^2} \frac{\cosh [(1-2x) \beta H ]}{\cosh \beta H} ,
\end{gathered}
\end{equation}
up to a term diminishing to zero as $M\to \infty$, 
where $H = \sqrt{m^2 + \Gamma^2}$ and $x$ is the shortest distance between $t/M$ and $\tpr/M$ on a ring. 
\rev{Note that the matrix consisting of elements given by \eqref{eq:Correlation} is circulant-symmetric, and thus its eigenvalues can be calculated explicitly using the Discrete Fourier Transform \cite{CirculantMatrices06} (see also \eqref{eq:FourierRep_q} and \eqref{eq:FourierRep_chi}). This offers }
\begin{align}\label{eq:Correlation_Eigenvalues}
\begin{split}
        \frac{\chi_k}{M}& = \delta_{k,0} \frac{m^2}{H^2} \sech^2 \beta H 
        + \frac{\beta\Gamma^2}{H} \frac{\tanh \beta H}{\beta^2 H^2 + \tilde{k}^2 \pi^2},
\end{split}
\end{align}
up to a term diminishing converging to zero as $M\to \infty$, 
where $\tilde{k} = \min(k, M-k)$.
\rev{See Appendix \ref{sec:eigenvalue_derivation} for a concise explanation and calculation. }
Therefore, for any bounded series $\{a_k\}_{k \geq 1}$,
the $k \geq 1$ modes of $\bm{\hq} = \eta \bm{\chi}(\bm{I}_M -  \eta \bm{\chi})^{-1}$ satisfy 
\begin{equation}
    \sum_{k = 1}^\infty a_k \frac{\eta \chi_k}{1 - \eta \chi_k} = O(1).
\end{equation}
The zeroth mode, on the other hand, is sufficiently large compared to the $k \geq 1$ modes, whose eigenvalue component is given by 
\begin{align}
        \frac{\eta \chi_0}{1 - \eta \chi_0 } &= \frac{\tanh \beta \Gamma + O(m^2)}{ \Gamma - \tanh \beta \Gamma + O(m^2) } \nonumber \\
        &= O\bigg({\rm max } \bigg\{ \frac{1}{m^2}, \frac{1}{\Delta \Gamma_\beta(\alpha) } \bigg\} \bigg).
\end{align}
This indicates that when studying the effective Hamiltonian \eqref{eq:Effective_Hamiltonian}, one only has to consider the zeroth mode of the 
long-range interaction $\bm{\hq} = \eta \bm{\chi}(\bm{I}_M -  \eta \bm{\chi})^{-1}$ when extracting its leading order behavior with respect to $\alpha$.
Under this simplification and qSA, it suffices to study the effective Hamiltonian given by
\begin{equation*}
    -H_{\rm eff,0} = \frac{\beta\hq_0}{2M^2} \qty( \sum_{t = 0}^{M-1} \sigma_t )^2 +  \sum_{t = 0}^{M-1} \big( \eta(\mathsf{z} + m) \sigma_t + B \sigma_t \sigma_{t+1} \big),
\end{equation*}
which is an Ising model whose partition function is analytically tractable using standard statistical mechanical techniques. 
Following the above observation, and rescaling $q_0 \leftarrow q_0/M$ and $\chi_0 \leftarrow \chi_0/M$, the equations of state reduce to the following set of equations, whose solution offers the correct behavior up to the leading order of $\alpha$:
\begin{align}\label{eq:hatq_SA}
    \hq_0 &= \frac{\alpha \beta \chi_0}{1 - \beta \chi_0}, \\
    \hchi_0 &= \frac{\alpha \beta (q_0 - \chi_0)}{(1-\beta \chi_0)^2} , \\
    m &=  \EE_{\mathsf{z}} \Bigg[ \Big\langle \frac{1}{M} \sum_{t = 0 }^{M-1} \sigma_t \Big \rangle _{\eff, 0} \Bigg] ,  \\
    \color{black}q_0 & \color{black}=  \EE_{\mathsf{z}} \Bigg[ \Big \langle \Big( \frac{1}{M} \sum_{t = 0}^{M-1} \sigma_t \Big)^2 \Big \rangle_{\eff, 0}\Bigg] , \\
    \label{eq:chi_SA}
    \color{black}\chi_0 &\color{black}= q_0 - \EE_{\mathsf{z}} \Bigg[ \Big\langle \frac{1}{M} \sum_{t = 0 }^{M-1} \sigma_t \Big \rangle ^2_{\eff, 0} \Bigg], 
\end{align}
where $\av{\cdots}_{\eff, 0}$ denotes the thermal average with respect to Hamiltonian $H_{\rm eff, 0}$ with temperature one. Under the limit $M\to \infty$, this analysis recovers the same fixed point as \cite{nishimoriQuantumEffectsNeural1996}  which is based on SA.
While SA is generally incorrect, it can accurately predict the critical point up to the leading order of $\alpha$ when following the lines of \cite{nishimoriQuantumEffectsNeural1996}. This result demonstrates the validity of SA in certain limiting cases, particularly when determining its critical behavior. 



\section{Numerical Solution for the Replica Symmetric Solution under Finite Trotter Size} \label{sec:numerics}
In this section, we report the numerical results obtained by solving the equations of state under qSA,  \eqref{eq:q_extr}--\eqref{eq:hatq_extr} and \eqref{eq:qSA_chihat}, which was done using successive substitution, or fixed point iteration. 

An obvious challenge when handling the equations of state is the intractable effective Hamiltonian \eqref{eq:Effective_Hamiltonian}, offering a difficulty in evaluating the moments \eqref{eq:q_extr} and \eqref{eq:chi_extr}. To obtain precise results even for large $M$, we employ the Fukui-Todo method \cite{fukuiOrderNClusterMonte2009}, which exploits the circulant property of the long-range interaction to perform highly efficient Markov Chain Monte Carlo (MCMC) sampling. On the other hand, we perform exact enumeration over all spin configurations for $M$ less than $19$. 
Finally, the average over $\mathsf{z}$ is performed using the Gauss-Hermite quadrature method consisting of 128 quadrature points. {Note that if qSA were not present, one would need to evaluate the average over a $M$-dimensional Gaussian distribution with covariance $\bm{\hchi}/\eta$, requiring further computational effort and therefore limit the extent of our numerical analysis. }

\begin{figure}
    \centering
    \includegraphics[width=0.75\linewidth]{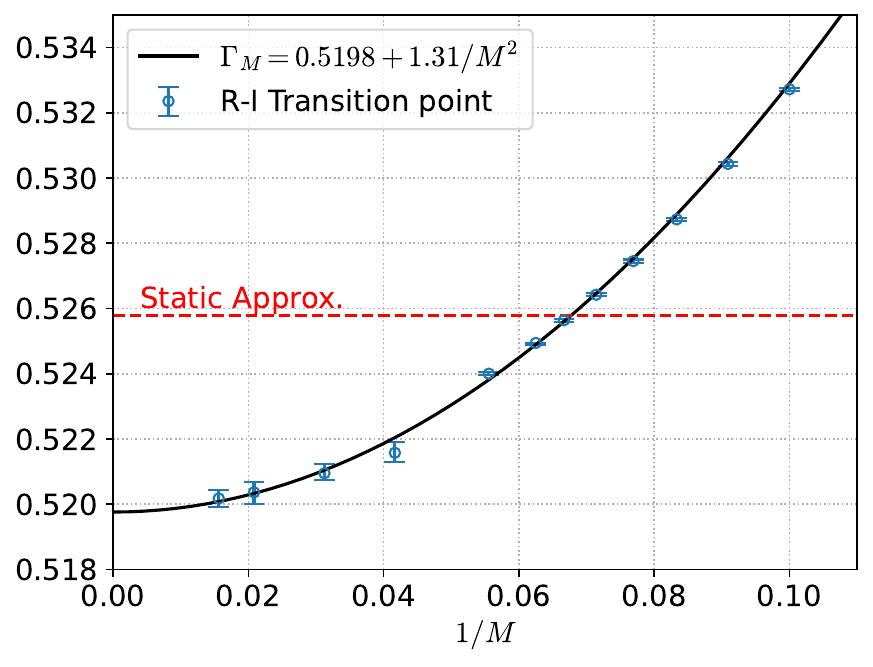}
    \caption{(Color online) The R-I transition point for different values of $M$. The black line corresponds to the extrapolated quadratic curve (whose first-order coefficient was found to be zero within error). }
    \label{fig:CritPoint_Mdependency}
\end{figure}

\begin{figure}
    \centering
    \includegraphics[width=0.75\linewidth]{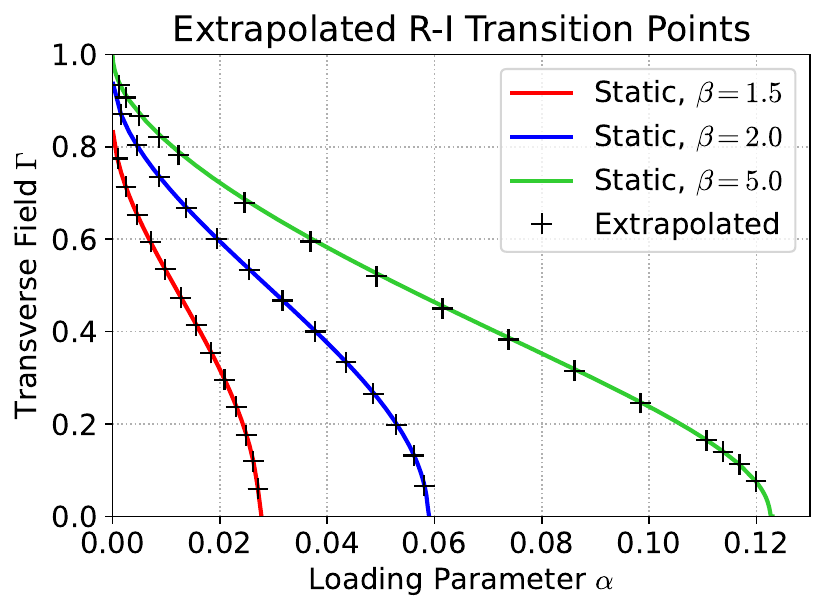}
    \caption{(Color online) The extrapolated transition point obtained by quadratic curve fitting on values from $M = 8,12,16$ and $32$ for various $\beta$.  }
    \label{fig:Extrapolated}
\end{figure}

\begin{figure*}
    \centering
    \includegraphics[width=0.85\linewidth]{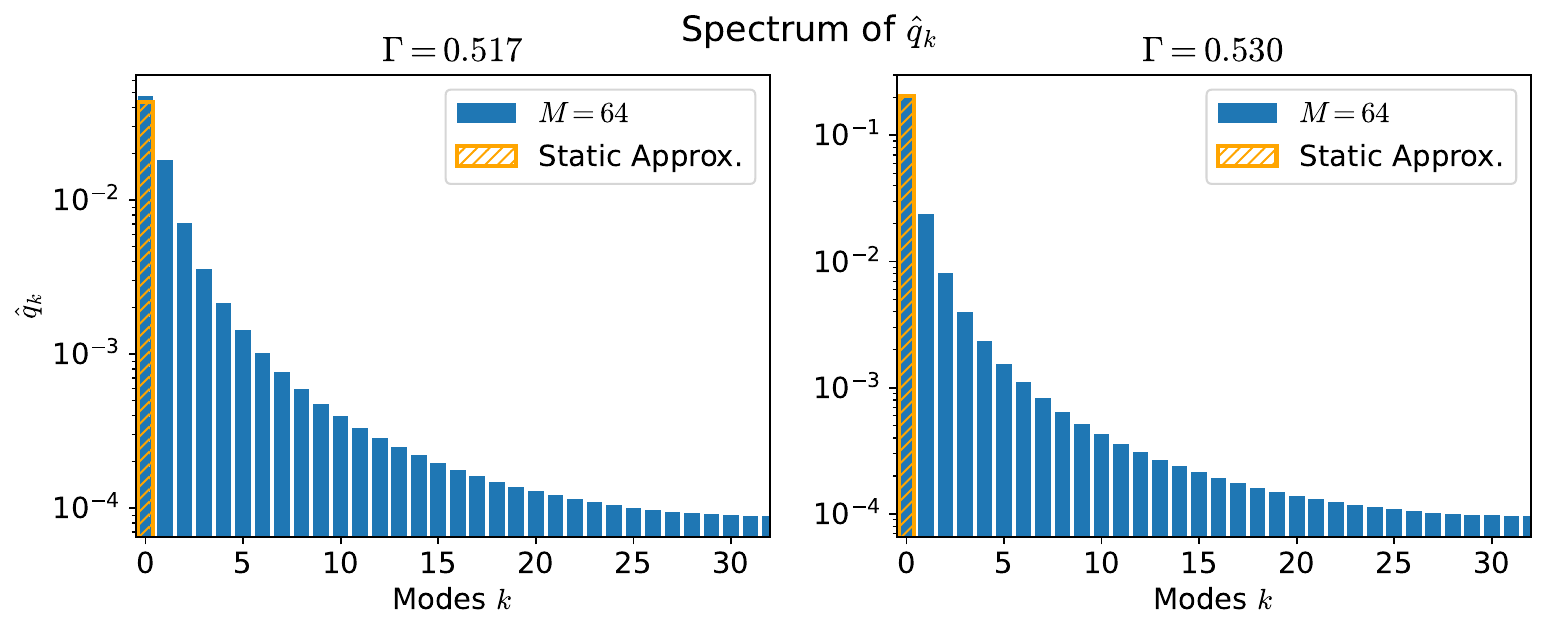}
    \caption{(Color online) The eigenvalues of $\bm{\hq}$ corresponding to each $k$-mode for the spin-glass state at $\Gamma = 0.530$, and the retrieval state at $\Gamma = 0.530$ under inverse temperature $\beta = 5$. These two points of $\Gamma$ correspond to the states just before and after the R-I transition (see figure \ref{fig:CritPoint_Mdependency}). }
    \label{fig:Spectrum}
\end{figure*}

In figure \ref{fig:PhaseDiagram_Beta5} we plot the phase diagram of the Hopfield model with respect to $\alpha$ and $\Gamma$ for $\beta = 5$ and different values of $M$. 
The R-I and R-II transition curves were obtained by performing the bisection method with respect to $\Gamma$ for each $\alpha$. 
Due to the need for precise free energy evaluation when obtaining the R-II transition curve, which is challenging with MCMC, the right-hand side of figure \ref{fig:PhaseDiagram_Beta5} only report the results for $M = 8$, $12$, and $16$, where exact enumeration remains feasible. 
Note also that we do not report the error bars for $M$ larger than $16$ in this figure, since the fluctuations resulting from different MCMC seeds were too small to be recognized. 
Notably, the R-I transition point shows minimal difference between SA and calculations performed with large Trotter size. 

To further investigate the discrepancy between these two, we perform precise numerical experiments for the case $\beta = 5$ and $\alpha = 0.0492$. In figure \ref{fig:CritPoint_Mdependency}, we plot the R-I transition point calculated from different values of $M$. For $M > 16$, we run 10 independent fixed point iterations with MCMC sampling using different random seeds and report the mean and standard error. Extrapolating the sequence of the estimated values to $M \to \infty$ via curve fitting to a quadratic function of $1/M$ with positive coefficients offers the estimate $\Gamma^\star_{\beta = 5}(\alpha = 0.0492) = 0.5198(5)$, which differs from $0.5257\cdots$ predicted by SA. 
Nevertheless, one can conclude that SA provides a qualitatively good approximation of the transition point when compared to the finite Trotter size approach considered in this paper. The same observations can be made for $\beta = 1.5$ and $2.0$ as well, which can be recognized from the extrapolated values obtained from $M = 8, 12,16$ and $32$ in figure \ref{fig:Extrapolated}. 

As explained in Section \ref{sec:RS_solution}, SA is essentially a rank-one approximation \eqref{eq:SA_qhat}, which projects the matrix $\bm{\hq}$ to the space spanned by $\bm{1}$. 
As it turns out, this approximation is valid even near the transition point. This is demonstrated in figure \ref{fig:Spectrum}, where we show each mode of $\bm{\hq}$ for $M = 64$, compared with $\hq_0$ in SA. It is evident that the modes decay fast with the mode number $k$, confirming the validity of a rank-one approximation. In fact, the value of $\hq_0$ from SA is very close to the eigenvalue corresponding to the zeroth mode in $\bm{\hq}$ for $M = 64$. This further indicates that the zeroth mode of susceptibility $\bm{\chi}$ under the effective model $\eqref{eq:Effective_Hamiltonian}$ 
is robust against the $k \geq 1$ modes appearing in the long-range interaction $\bm{\hq}$, hence discarding them all together (or equivalently, projecting it to $\bm{1}$ space) may lead to a reasonable approximation. 

\section{Conclusion}
In this work, we presented the replica symmetric analysis of the Hopfield model under a uniform transverse field. Unlike the analysis based on the static approximation, which ignores the imaginary time dependency of the order parameters for the sake of taking the Trotter size to infinity, we instead handle the imaginary time dependencies exactly by keeping the Trotter size finite. The analysis indicates that while the static approximation is valid near $\alpha = 0$, a weaker version of this approximation, termed the quasi-static ansatz, holds within the replica symmetry. In fact, the static approximation can be interpreted as further imposing a rank-one approximation on top of the quasi-static ansatz.
Numerical experiments indicate a quantitative difference between results obtained from the static approximation and the approach considered in this paper. It should be stressed that this difference is minute; evidence on the validity of the rank-one approximation induced by the static approximation is also provided numerically. 
Whether this observation also holds for systems under different quantum effects, such as antiferromagnetic transverse interactions \cite{Seki12, sekiQuantumAnnealingAntiferromagnetic2015} is an interesting direction for future work. 

\section*{Acknowledgements} \label{sec:acknowledgements}
    The authors would like to thank Tomoyuki Obuchi, Yoshinori Hara and Takashi Takahashi for insightful discussions. This work was supported by JSPS KAKENHI Grant No. 22KJ1074 (KO), JST CREST Grant Number JPMJCR1912 (YK), and Grant-in-Aid for Transformative Research Areas (A), “Foundation of Machine Learning Physics” (22H05117) (YK). 

\appendix
\begin{widetext}
\section{Detailed Calculation of the Replica Symmetric Free Energy}
\label{sec:RS_FE_derivation}
\subsection{Calculation of the average replicated partition function}
The objective of this section is to calculate the logarithm of the partition function \eqref{eq:Trotter_Partition} averaged over the disorder $\{\bm{\xi}^\mu\}_{\mu =1}^P$. The $n$-replicated partition function averaged over the quenched randomness is given by 
\begin{equation}\label{eq:replicated_partition_function}
     \EE Z_M^n =  \EE \mathop{\Tr}_{\{\sigma^a_{i,t}\}  } \exp \sum_{a = 1}^n \sum_{t=0}^{M-1}  \qty[  \frac{\beta}{2NM} \sum_{i,j} \sigma_{i,t}^a \sigma_{j,t}^a \qty( \xi_i^1 \xi_j^1 + \sum_{\mu = 2}^P \xi_i^\mu \xi_j^\mu) + B \sum_{i = 1}^N \sigma_{i,t}^a \sigma_{i, t+1}^a ]. 
\end{equation}

We consider the case where the spins are aligned with the first pattern, which we took as $\bm{\xi}^1 = \bm{1}$ for the sake of convenience. Therefrom, define the order parameters 
\begin{align}\label{eq:replicated_order_parameter}
    m_t^a \equiv \frac{1}{N} \sum_{i = 1}^N \sigma_{i,t}^a, \quad Q_{t\tpr}^{ab} \equiv \frac{1}{N} \sum_{i = 1}^N \sigma_{i,t}^a \sigma_{i,\tpr}^b, 
\end{align}
for $t, \tpr = 0,\ldots,M-1$ and $1 \leq a \leq b \leq n$. Conditioned on a fixed configuration of $\{\sigma^a_{i,t}\}$, the local field $
h_{t,\mu}^a \equiv \sum_{i = 1}^N \sigma_{i,t}^a \xi_i^\mu / \sqrt{N}
$
for $t = 0, \ldots, M-1$ and $\mu = 2,\ldots, P$ is a centered Gaussian random variable from the central limit theorem. 
Its covariance is given by 
\begin{equation}
    \EE h_{t,\mu}^a h_{\tpr,\nu}^b = \delta_{\mu\nu} Q_{t\tpr}^{ab}. 
\end{equation}
Noticing that the averaged replicated partition function \eqref{eq:replicated_partition_function} is only dependent on the quenched randomness via $\{h_{t,\mu}^a\}$, and by introducing delta functions to enforce the constraint \eqref{eq:replicated_order_parameter}, \eqref{eq:replicated_partition_function} can be recast as 
\begin{equation}
\begin{gathered}
         \mathop{\Tr}_{\{\sigma^a_{i,t}\}  } \int \prod_{a\leq b} \prod_{t,\tpr} \dd Q_{t\tpr}^{ab} \delta \qty(  N Q_{t\tpr}^{ab} - \sum_{i = 1}^N \sigma_{i,t}^a \sigma_{i,t}^b ) \prod_{a = 1}^n \prod_{t=0}^{M-1} \delta \qty( N m_t^a - \sum_{i = 1}^N \sigma_{i,t}^a ) \\
    \times \exp \sum_{a = 1}^n \sum_{t=0}^{M-1} \qty[   \frac{N \eta }{2} (m_t^a)^2 + B \sum_{i = 1}^N \sigma_{i,t}^a \sigma_{i,t+1}^a ]
    \qty{\EE_{\bm{h} | \vb{Q} } \exp \qty( \frac{\eta}{2} \sum_{a = 1}^n \sum_{t=0}^{M-1} (h_t^a )^2 ) }^{P-1},
\end{gathered}
\end{equation}
where $\EE_{\bm{h} | \vb{Q}} $ denotes the average with respect to $\{ h_t^a \}$, whose law is equivalent to that of $\{h_{t,\mu}^a\}$ for any $\mu = 2,\ldots , P$. 
Further decoupling the delta functions using its Fourier representation, \textit{i.e.} $\delta(NQ_{t\tpr}^{ab} - \cdots) = \int_\mathbb{C} \frac{\eta \dd \hq_{t\tpr}^{ab}  }{2\pi } e^{ \eta \hQ_{t\tpr}^{ab} (N Q_{t\tpr}^{ab} - \cdots)} $ and $\delta (N m_t^a - \cdots) =  \int_\mathbb{C} \frac{\eta \dd \ham_{t}^{a}  }{2\pi} e^{ -\eta \ham_{t}^{a} ( N m_{t}^{a} - \cdots)} $ and using the differential operator expression for the average $\EE_{\bm{h} | \vb{Q}} $, the averaged replicated partition function takes the following form up to leading order of $N$: 
\begin{align}
\begin{split}
           \EE Z_M^n = & \int \prod_{a \leq b} \prod_{t, \tpr} \dd Q_{t\tpr}^{ab} \dd \hQ_{t\tpr}^{ab} \prod_{a = 1}^n \prod_{t=0}^{M-1} \dd m_t^a \dd \ham_t^a \exp N \eta \Bigg[ \frac{1}{2} \sum_{t,\tpr} \sum_{a, b} Q_{t\tpr}^{ab} \hQ_{t\tpr}^{ab} + \frac{1}{2} \sum_{t=0}^{M-1} \sum_{a = 1}^n(  (m_t^a)^2 - 2m_t^a \ham_t^a) \Bigg]  \\
       & \times \exp \qty[  N \log \mathop{\Tr}_{\{ \sigma_{t}^a \}} e^\mathcal{L} + P \mathcal{E} ],
\end{split}
\end{align}
where 
\begin{gather}
    \mathcal{L}  = -\frac{\eta}{2} \sum_{a, b} \sum_{t,\tpr} \hQ_{t\tpr}^{ab} \sigma_t^a \sigma_\tpr^b + \sum_{a= 1}^n \sum_{t=0}^{M-1} ( \eta \ham_{t}^a \sigma_t^a + B \sigma_t^a \sigma_{t + 1}^a ), \\
\label{eq:energetic_term_replicated}
    \mathcal{E} = \log \EE_{\bm{h} | \vb{Q} } \exp \qty( \frac{\eta}{2} \sum_{a = 1}^n \sum_{t=0}^{M-1} (h_t^a )^2 ) = \log \qty[ \eval{ \exp \qty( \frac{1}{2}\sum_{a,b}\sum_{t,\tpr} Q_{t\tpr}^{ab} \pdv{}{h_t^a}{h_\tpr^b} ) \exp \qty(\frac{\eta}{2} \sum_{a = 1}^n \sum_{t=0}^{M-1} (h_t^a )^2 )}_{\bm{h} = \bm{0}}],
\end{gather}
 and we defined $Q_{\tpr t}^{ba} = Q_{t\tpr}^{ab}$ for $a \leq b$. 
Utilizing the saddle-point method, the logarithm of the average replicated partition function can finally be expressed as an extremum problem over the order parameters: 
\begin{align}\label{eq:replicated_partition_function_ansatz_free}
   \lim_{N \to \infty} \frac{1}{nN} \log \EE Z_M^n = \frac{1}{n} \mathop{\rm Extr} \Bigg\{ &\frac{\eta}{2} \sum_{t,\tpr}\sum_{a, b} Q_{t\tpr}^{ab}\hQ_{t\tpr}^{ab} +\frac{\eta}{2} \sum_{t=0}^{M-1} \sum_{a = 1}^n(  (m_t^a)^2 - 2m_t^a \ham_t^a) + \log \mathop{\Tr}_{\{ \sigma_{t}^a \}} e^\mathcal{L} + \alpha \mathcal{E} \Bigg\} .
\end{align}
\end{widetext}

The extremum conditions are given by the following equations:
\begin{align}
    Q_{t\tpr}^{ab}  &= \frac{ \Tr \sigma_{t}^a \sigma_\tpr^b e^\mathcal{L} }{ \Tr e^\mathcal{L}}, \\
    m_t^a &= \ham_t^a = \frac{ \Tr \sigma_{t}^a e^\mathcal{L} }{ \Tr e^\mathcal{L}}, \\
    \hQ_{t\tpr}^{ab} &= e^{-\mathcal{E}} \EE_{\bm{h} | \vb{Q} } \qty[ \pdv{}{h_t^a}{h_\tpr^b} e^{\frac{\eta}{2} \sum_{a ,t} (h_t^a)^2 } ].
\end{align}

\subsection{Calculation under replica symmetry} 
To obtain a formula for \eqref{eq:replicated_partition_function_ansatz_free} whose $n\to 0$ limit can be taken analytically, we consider the replica symmetric ansatz: 
\begin{align}
    Q_{t\tpr}^{ab} &= \begin{cases}
        q_{t\tpr}  & a = b\\
        q_{t\tpr} - \chi_{t\tpr}& a \neq b
    \end{cases}, \\
    \hQ_{t\tpr}^{ab} &= \begin{cases}
       -\hq_{t\tpr} -\hchi_{t\tpr} & a = b \\ 
         - \hchi_{t\tpr}  & a \neq b
    \end{cases}, \\
    m_t^a  &(= \hat{m}_t^a) = m_t. 
\end{align}
This assumption yields the following equalities :
\begin{align}
\label{eq:doubledoubledouble}
   \frac{1}{n }& \sum_{a, b} Q_{t\tpr}^{ab}\hQ_{t\tpr}^{ab} = -\chi_{t\tpr} \hchi_{t\tpr} - q_{t\tpr} \hq_{t\tpr} + O(n), \\
   \mathcal{L} &= \sum_{a = 1}^n \qty( \sum_{t,\tpr} \frac{ \eta \hq_{t\tpr}}{2} \sigma_t^a \sigma_\tpr^a +  \sum_{t=0}^{M-1} (\eta m_t \sigma_t^a + B \sigma_t^a \sigma_{t +1}^a) ) \nonumber \\ \label{eq:altered_L}
   & \qquad \qquad \qquad + \frac{\eta \hchi_{t\tpr} }{2} \qty( \sum_{a = 1}^n \sigma_t^a ) \qty( \sum_{a = 1}^n \sigma_\tpr^a ).
\end{align}
Using the Hubbard-Stratonovich transformation on the last term in \eqref{eq:altered_L}, one can immediately see that 
\begin{equation}
\label{eq:L_simplified}
     \mathop{\Tr}_{\{ \sigma_{t}^a \}} e^\mathcal{L} = \EE_{\bm{z} \sim \mathcal{N}(\bm{0}, \bm{\hchi} / \eta)} \qty( \mathop{\Tr}_{\{ \sigma_{t} \}} e^{-H_{\rm eff}} )^n .
\end{equation}
On the other hand, using the shorthand notation for the $n$-dimensional nabla $(\pdv{}{h_t^1}, \ldots, \pdv{}{h_t^n} )^\ten = \nabla_t $, the differential operator in \eqref{eq:energetic_term_replicated} simplifies to 
    \begin{align}
&\exp \qty( \frac{1}{2}\sum_{a,b}\sum_{t,\tpr} Q_{t\tpr}^{ab} \pdv{}{h_t^a}{h_\tpr^b} ) \nonumber \\
    = &\exp \sum_{t,\tpr} \qty[  \frac{\chi_{t\tpr}}{2} \nabla_t {\cdot} \nabla_\tpr + \frac{q_{t\tpr} - \chi_{t\tpr}}{2} \qty( \nabla_t^\ten \bm{1} ) \qty( \nabla_\tpr^\ten \bm{1} ) ] \nonumber\\ 
    = &\EE_{\bm{\Xi} \sim \mathcal{N} ( \bm{0}, \bm{q} - \bm{\chi} )} \exp \qty[ \sum_{t,\tpr}  \frac{\chi_{t\tpr}}{2} \nabla_t {\cdot} \nabla_\tpr + \sum_{t=0}^{M-1} \hspace{-0.16667em} \Xi_t  \qty( \nabla_t^\ten \bm{1} ) ],
\end{align}
where we again used the Hubbard-Stratonovich transformation to go from the first line to the second line. Note that $e^{a \partial_x} f(x) = f(x+a)$, \textit{i.e.} the last term in the above expression is a translational operator. Under this observation, 
\begin{align}
    \frac{\mathcal{E}}{n} &= \frac{1}{n} \log \qty[ \EE_{\bm{\Xi} \sim \mathcal{N} (\bm{0}, \bm{q} - \bm{\chi})} \qty( \EE_{\bm{h} \sim \mathcal{N}(\bm{0}, \bm{\chi} )}e^{\frac{\eta}{2} \norm{\bm{h} + \bm{\Xi}}_2^2 } )^n] \nonumber \\
    &=  \EE_{\bm{\Xi} \sim \mathcal{N} (\bm{0}, \bm{q} - \bm{\chi})} \log \EE_{\bm{h} \sim \mathcal{N}(\bm{0}, \bm{\chi} )}e^{\frac{\eta}{2} \norm{\bm{h} + \bm{\Xi}}_2^2 } +O(n) \nonumber \\
    &=\frac{1}{2} \log \det (\bm{I}_M - \eta \bm{\chi}) \nonumber \\
    \label{eq:E_simplified}
    & \qquad - \frac{\eta}{2} \Tr (\bm{I}_M - \eta \bm{\chi})^{-1} (\bm{q} - \bm{\chi}) + O(n).
\end{align}
Accumulating \eqref{eq:replicated_partition_function_ansatz_free}, \eqref{eq:doubledoubledouble}, \eqref{eq:L_simplified} and \eqref{eq:E_simplified} yields the main result \eqref{eq:replica_symmetric_free_energy}.

\section{Detailed Calculation of Equation \eqref{eq:Correlation_Eigenvalues}}
\label{sec:eigenvalue_derivation}
Here, we present a detailed calculation of the eigenvalues of the matrix given by \eqref{eq:Correlation}, which we 
repeat here for sake of convenience: 
\begin{equation}
\begin{gathered}
      \av{\sigma_t \sigma_\tpr} - \av{\sigma_t} \av{\sigma_\tpr}  \\ =\frac{m^2}{H^2} \sech^2\beta H + \frac{\Gamma^2}{H^2} \frac{\cosh [(1-2x) \beta H ]}{\cosh \beta H} ,
\end{gathered}
\end{equation} 
where $H = \sqrt{m^2 + \Gamma^2}$ and $x$ is the shortest distance between $t/M$ and $\tpr/M$ on a ring. 
For a general symmetric circulant matrix of the form $\vb{K}_{t\tpr} = (K_{\abs{t-\tpr}})$, with $\abs{\cdot}$ measuring the shortest distance between $t$ and $\tpr$ on a ring, its eigenvalues are given by \cite{CirculantMatrices06}
\begin{equation}\label{eq:cosine_summation}
    \lambda_k = \sum_{L = 0}^{M-1} K_{\min(L, M - L )} \cos\qty( \frac{2\pi kL }{M})
\end{equation}
for $k = 0, \cdots, M-1$. From the property of the cosine function, we have that $\cos\qty( \frac{2\pi kL }{M}) = \cos\qty( \frac{2\pi (M-k)L }{M}) = \cos\qty( \frac{2\pi k(M-L) }{M})$. As a consequence, it suffices to only consider the case $k \leq M/2$ since $\lambda_k = \lambda_{M-k}$. Additionally, the sum over $L$ running from $0$ to $M-1$, can also be confined from $0$ to $\lfloor M/2 \rfloor$ up to an $O(1)$ term correction. Combining the above observations, and substituting the explicit expression for $\av{\sigma_t \sigma_\tpr} - \av{\sigma_t} \av{\sigma_\tpr}$ into \eqref{eq:cosine_summation} yields 
\begin{widetext}
\begin{equation}\label{eq:chi_n_intermediate}
    M \chi_k = 2\sum_{L = 0}^{\lfloor M/2 -1 \rfloor}  \qty( \frac{m^2}{H^2} \sech^2\beta H + \frac{\Gamma^2}{H^2} \frac{\cosh [(1-2L/M) \beta H ]}{\cosh \beta H} ) \cos\qty( \frac{2\pi kL }{M}) + O(1),  
\end{equation}
where the last $O(1)$ term is equal to the term in the above summand with $L = (M+1)/2$ if $M$ is odd, and $0$ if $M$ is even. In any case, it is bounded by a constant from above for any choice of $M$ and finite $\beta, m$ and $\Gamma$. 

The first term in \eqref{eq:chi_n_intermediate} is non-zero if and only of $n = 0$. The second term can be expressed as an integral in the large $M$ limit, yielding 
\begin{equation}
  \chi_k = \delta_{k,0} \frac{m^2}{H^2} \sech^2 \beta H +  \frac{2\Gamma^2}{H^2} \int_0^{\frac{1}{2}} \dd x \frac{\cosh[(1-2x)\beta H]}{\cosh \beta H} \cos(2\pi k x) + o_M(1),
\end{equation}
where $o_M(1)$ denotes a term converging to $0$ in the limit $M\to \infty$. Performing the integral, and noting that $\chi_k = \chi_{M-k}$ once again yields \eqref{eq:Correlation_Eigenvalues}. 
\end{widetext}

\bibliography{ref}

\end{document}